\documentclass[reprint,aip,jcp]{revtex4-2}%
\usepackage{amsfonts}
\usepackage{amsmath}
\usepackage{amssymb}
\usepackage{graphicx}
\usepackage{dcolumn}
\usepackage{bm}
\usepackage{float}
\providecommand{\U}[1]{\protect\rule{.1in}{.1in}}
\begin{document}
\title{Finite-temperature vibronic spectra from the split-operator coherence
thermofield dynamics}
\author{Zhan Tong Zhang}
\affiliation{Laboratory of Theoretical Physical Chemistry, Institut des Sciences et
Ing\'enierie Chimiques, Ecole Polytechnique F\'ed\'erale de Lausanne (EPFL),
CH-1015 Lausanne, Switzerland}
\author{Ji\v{r}\'{\i} J. L. Van\'{\i}\v{c}ek}
\email{jiri.vanicek@epfl.ch}
\affiliation{Laboratory of Theoretical Physical Chemistry, Institut des Sciences et
Ing\'enierie Chimiques, Ecole Polytechnique F\'ed\'erale de Lausanne (EPFL),
CH-1015 Lausanne, Switzerland}
\date{\today}

\begin{abstract}
We present a numerically exact approach for evaluating vibrationally resolved
electronic spectra at finite temperatures using the coherence thermofield
dynamics. In this method, which avoids implementing an algorithm for solving
the von Neumann equation for coherence, the thermal vibrational ensemble
is first mapped to a pure-state wavepacket in an augmented space, and this
wavepacket is then propagated by solving the standard, zero-temperature
Schr\"{o}dinger equation with the split-operator Fourier method. We show that
the finite-temperature spectra obtained with the coherence thermofield
dynamics in a Morse potential agree exactly with those computed by
Boltzmann-averaging the spectra of individual vibrational levels. Because the
split-operator thermofield dynamics on a full tensor-product grid is
restricted to low-dimensional systems, we briefly discuss how the accessible
dimensionality can be increased by various techniques developed for the
zero-temperature split-operator Fourier method.

\end{abstract}
\maketitle

\graphicspath{{./Figures/}{../Figures/}{"C:/Users/GROUP
LCPT/Documents/Tong/tfd_qm/Figures/"}
{C:/Users/Jiri/Dropbox/Papers/Chemistry_papers/2023/Exact_quantum_TFD/Figures/}}

\section{Introduction}

Vibrationally resolved electronic spectroscopy has proven to be a powerful
tool for probing molecular systems and understanding photochemical processes.
These experiments are typically conducted at ambient or higher temperatures,
which are relevant to understanding biochemical and atmospheric processes as
well as developing new
materials.\cite{Peterman_vanAmerongen:1997,Gaiduk_Orrit:2010,Qian_Chen:2020,Ndengue_Osborn:2023}
Other than the appearance of hot bands and the thermal Doppler broadening
effects, non-Condon effects can also noticeably alter spectral features at
finite
temperatures.\cite{Orlandi:1973,Tanimura_Mukamel:1993,Kundu_Makri:2022,Roy_Fleming:2022}
These thermal effects may become more pronounced in two-dimensional
spectroscopy experiments.\cite{Begusic_Vanicek:2021a} Including temperature
effects in theoretical simulations is thus essential to accurately interpret
and predict the results of vibronic
spectroscopy.\cite{Zucchelli_Cremonesi:2002,Baiardi_Barone:2013,Qian_Chen:2020,Begusic_Vanicek:2020,Xun_He:2020,Srsen_Heger:2020,Wernbacher_Gonzalez:2021,Begusic_Vanicek:2021a,Prlj_Curchod:2022,Ndengue_Osborn:2023}%

Given the large number of vibrational levels that may be occupied,
time-independent sum-over-state expressions become impractical at higher
temperatures for more than a few vibrational degrees of freedom.
Time-dependent approaches offer a promising alternative, in which the spectrum
is evaluated as the Fourier transform of an appropriate autocorrelation
function without the need to select relevant transitions and compute their
Franck--Condon factors.\cite{Heller:1981a,Baiardi_Barone:2013} However, the
most widely used quantum
mechanical\cite{Feit_Steiger:1982,Meyer_Cederbaum:1990,Ben-Nun_Martinez:1999,Worth_Burghardt:2004,book_Tannor:2007}
and
semiclassical\cite{Heller:1981,Heller:1981a,Herman_Kluk:1984,Hagedorn:1998,book_Lubich:2008,book_Heller:2018,Bonfanti_Pollak:2018}
methods were developed only for pure states and cannot directly handle a
thermal ensemble of states.

Standard approaches to include temperature effects either directly evolve the
thermal density matrix or employ statistical averaging over pure states. An
alternative way to represent quantum dynamics at finite temperature, called
thermofield dynamics,\cite{Suzuki:1985,Takahashi_Umezawa:1996} maps the
thermal ensemble to a pure state in an augmented space, which can be
propagated by conventional wavepacket methods for solving the time-dependent
Schr\"{o}dinger equation. Thermofield dynamics has been applied to problems in
electronic structure\cite{Harsha_Scuseria:2019} and molecular quantum
dynamics,\cite{Borrelli_Gelin:2016,Borrelli_Gelin:2017,Gelin_Borrelli:2017,Chen_Zhao:2017,Borrelli_Gelin:2021,Fischer_Saalfrank:2021,Gelin_Borrelli:2023}
including the simulation of vibronic
spectra.\cite{Reddy_Prasad:2015,Reddy_Prasad:2016,Begusic_Vanicek:2020,Begusic_Vanicek:2021a,Begusic_Vanicek:2021b,Chen_Gelin:2021,Polley_Loring:2022}
In a similar spirit, Grossmann performed exact calculations of the
thermal dipole time correlation function for vibrational spectra by
propagating two thermal wavepackets and evaluating the matrix element of the
dipole moment between them.\cite{Grossmann:2014}

{In vibronic spectroscopy applications, the evolution of the thermofield
wavepacket replaces the dynamics of the coherence between two electronic
states. Due to the doubling of dimensions, the coherence thermofield dynamics
obviously benefits from a combination with semiclassical or other approximate
propagation methods that scale favorably with the number of degrees of
freedom. Combining the thermofield dynamics with the extended thawed Gaussian
approximation\cite{Lee_Heller:1982,Patoz_Vanicek:2018,Prlj_Vanicek:2020} has
enabled the simultaneous description of finite-temperature,
non-Condon,\cite{Reddy_Prasad:2016} and even anharmonicity
effects.\cite{Begusic_Vanicek:2020,Begusic_Vanicek:2021a,Begusic_Vanicek:2021b}
Yet, one would sometimes like to avoid dynamical approximations and obtain the
exact quantum solution. }

For small systems with a known potential energy surface, the dynamical Fourier
method may be used to solve the time-dependent Schr\"{o}dinger equation on a
grid with the split-operator
algorithm.\cite{Feit_Steiger:1982,Kosloff_Kosloff:1983,Kosloff_Kosloff:1983a,book_Tannor:2007,book_Lubich:2008}
Although limited to low dimensions, the split-operator Fourier method offers a
numerically exact solution, which can validate semiclassical and other more
approximate approaches.

Here, we combine the coherence thermofield dynamics with the split-operator
Fourier method in order to simulate vibronic spectra at finite temperature.
This approach is validated on a one-dimensional anharmonic Morse model by
comparing the spectra computed with the split-operator coherence thermofield
dynamics to the results of the Boltzmann averaging of spectral contributions
from different initial vibrational levels. We also examine the thermofield
wavepacket's evolution in the augmented configuration space consisting of both
the physical and fictitious degrees of freedom.

\section{Theory}

Let us consider a molecular system with $D$ nuclear degrees of freedom and two
uncoupled adiabatic electronic states: the ground state $|g\rangle$ and the
excited state $|e\rangle$. The absorption spectrum for vibronic
transitions from $|g\rangle$ to $|e\rangle$ is obtained as the Fourier
transform,
\begin{equation}
\sigma(\omega)=\operatorname{const}\cdot\int_{-\infty}^{\infty}C(t)e^{i\omega
t}\,dt, \label{eq:spec}%
\end{equation}
of the dipole time correlation function,
\begin{equation}
C(t)=\operatorname{Tr}(\hat{\mu}^{\dagger}e^{-i\hat{H}_{e}t/\hbar}\hat{\mu
}\hat{\rho}e^{i\hat{H}_{g}t/\hbar}), \label{eq:thermal_ct}%
\end{equation}
where $\hat{\mu}$ represents the transition dipole moment operator, $\hat
{H}_{g}$ and $\hat{H}_{e}$ are the vibrational Hamiltonians associated with
the ground and excited electronic states, $\hat{\rho}$ is the vibrational
density operator of the ground electronic state, and
$\operatorname{const}=2\pi\omega/(\hbar c)$%
.\cite{Heller:1981a,book_Tannor:2007} As we are interested only in the
dependence of the spectrum on frequency and temperature, in the following we
set $\operatorname{const}=1/(2*\pi)$.

At a finite temperature $T$, the density of occupied vibrational levels in the
initial population follows the Boltzmann distribution
\begin{equation}
\hat{\rho}=e^{-\beta\hat{H}_{g}}/\operatorname{Tr}(e^{-\beta\hat{H}_{g}}),
\end{equation}
with the inverse temperature $\beta=1/(k_{B}T)$. We can then rewrite $C(t)$
as
\begin{equation}
C(t)=\sum_{n}p_{n}e^{i\omega_{n,g}t}\langle n,g|\hat{\mu}^{\dagger}%
e^{-i\hat{H}_{e}t/\hbar}\hat{\mu}|n,g\rangle, \label{eq:boltz}%
\end{equation}
where $p_{n}=e^{-\beta\hbar\omega_{n,g}}/\sum_{k}e^{-\beta\hbar\omega_{k,g}}$
is the Boltzmann probability, $|n,g\rangle$ denotes the $n$-th vibrational
eigenfunction of $\hat{H}_{g}$, and $\hbar\omega_{n,g}$ is the corresponding vibrational energy.\cite{Baiardi_Barone:2013} The thermal spectrum can be
obtained from
\begin{equation}
C(t)=\sum_{n}p_{n}e^{i\omega_{n,g}t}\langle\varphi_{n}(0)|\varphi
_{n}(t)\rangle
\end{equation}
by propagating the initial vibrational wavepackets $|\varphi_{n}%
(0)\rangle=\hat{\mu}|n,g\rangle$ individually as $|\varphi_{n}(t)\rangle
=e^{-i\hat{H}_{e}t/\hbar}|\varphi_{n}(0)\rangle$. In high-dimensional systems,
propagating wavepackets associated with all occupied vibrational levels
becomes impractical, and a statistical sampling approach is
needed.\cite{Manthe_Larranaga:2001,book_MCTDH,Barbatti_Sen:2016,Wernbacher_Gonzalez:2021,Prlj_Curchod:2022}%

Using the coherence thermofield dynamics, the thermal ensemble is instead
mapped to a pure state in a doubled configuration space. The thermal dipole
time correlation function (\ref{eq:thermal_ct}) can be rewritten
exactly\cite{Begusic_Vanicek:2020} (see the Appendix) as a
wavepacket autocorrelation function,
\begin{equation}
C(t)=\langle\bar{\psi}|e^{-i\hat{\bar{H}}t/\hbar}|\bar{\psi}\rangle,
\label{eq:tfd_ct}%
\end{equation}
of an initial thermofield wavepacket, represented in the augmented thermofield
space (denoted by a bar) as
\begin{equation}
\bar{\psi}(\bar{q})=\langle q|\hat{\mu}\hat{\rho}^{1/2}|q^{\prime}\rangle,
\end{equation}
where $\bar{q}=(q,q^{\prime})^{T}$ is a $2D$-dimensional coordinate vector in
the augmented\ configuration space and $|q^{\prime}\rangle$ represents the
position states in the \textquotedblleft fictitious\textquotedblright\ Hilbert
space (denoted by a prime). The wavepacket is propagated according to the
time-dependent Schr\"{o}dinger equation with the \textquotedblleft
augmented\textquotedblright\ Hamiltonian $\hat{\bar{H}}$ expressed in the
position representation as
\begin{equation}
\bar{H}(\bar{q})=H_{e}(q)-H_{g}(q^{\prime}), \label{eq:tfd_hamil}%
\end{equation}
with a $2D\times2D$ mass matrix,
\begin{equation}
\bar{m}=%
\begin{pmatrix}
m & 0\\
0 & -m
\end{pmatrix},
\end{equation}
replacing the $D\times D$ mass matrix $m$. {In contrast to standard
thermofield dynamics, where the two Hamiltonians in the right-hand side of
Eq.~(\ref{eq:tfd_hamil}) are identical, the thermofield Hamiltonian~(\ref{eq:tfd_hamil}) in vibronic spectroscopy depends on the two Hamiltonians
associated with the two electronic states involved in the transition. In
essence, the thermofield wavepacket encodes the coherence between the dynamics
on the two potential energy surfaces.}

If the ground surface
\begin{equation}
V_{g}(q^{\prime})=v_{0}+(q^{\prime}-q_{\text{ref}}^{\prime})^{T}\cdot
\kappa\cdot(q^{\prime}-q_{\text{ref}}^{\prime})/2\label{eq:harmonic}%
\end{equation}
is harmonic, the initial thermofield wavepacket\cite{Begusic_Vanicek:2020}
\begin{equation}
\psi(\bar{q})=e^{(i/\hbar)[(\bar{q}-\bar{q}_{0})^{T}\cdot\bar{A}_{0}\cdot
(\bar{q}-\bar{q}_{0})/2+\bar{p}_{0}^{T}\cdot(\bar{q}-\bar{q}_{0})+\bar{\gamma
}_{0}]}%
\end{equation}
is a Gaussian centered at
\begin{equation}
\bar{q}_{0}=%
\begin{pmatrix}
q_{\text{ref}}^{\prime}\\
q_{\text{ref}}^{\prime}%
\end{pmatrix}
,\qquad\bar{p}_{0}=%
\begin{pmatrix}
0\\
0
\end{pmatrix}
\end{equation}
in the phase space. This Gaussian has a temperature-dependent width
matrix,\cite{Begusic_Vanicek:2020}
\begin{equation}
\bar{A}_{0}=i%
\begin{pmatrix}
\mathrm{A} & \mathrm{B}\\
\mathrm{B} & \mathrm{A}%
\end{pmatrix}
,\label{eq:tfd_width}%
\end{equation}
where $\mathrm{A}$ and $\mathrm{B}$ are the $D\times D$ block matrices
\begin{align}
&  \mathrm{A}=m^{1/2}\cdot\Omega\cdot\operatorname{coth}(\beta\hbar
\Omega/2)\cdot m^{1/2},\label{eq:block_A}\\
&  \mathrm{B}=-m^{1/2}\cdot\Omega\cdot\sinh(\beta\hbar\Omega/2)^{-1}\cdot
m^{1/2},\label{eq:block_B}%
\end{align}
with $\Omega=\sqrt{m^{-1/2}\cdot\kappa\cdot m^{-1/2}}$. The phase factor,
\begin{equation}
\bar{\gamma}_{0}=(-i\hbar/2)\ln[\operatorname{det}(m\cdot\Omega/\pi\hbar)],
\end{equation}
serves to ensure normalization.

The spectrum at a finite temperature can now be obtained by applying standard
techniques for solving the time-dependent Schr\"{o}dinger equation to the
thermofield wavepacket with a doubled number of coordinates. {Although the
increase in dimensionality implies that an exact thermofield solution does not
save computational time compared to the direct propagation of the density
matrix, the wavepacket approach permits the use of zero-temperature numerical
methods and avoids implementing methods for solving the von Neumann equation.
}

Among various numerical methods for propagating wavepackets, we choose to
combine the thermofield dynamics with the second-order split-operator
algorithm,\cite{Feit_Steiger:1982,Kosloff_Kosloff:1983,Kosloff_Kosloff:1983a,book_Tannor:2007,book_Lubich:2008}%
 a popular choice for separable Hamiltonians, because it is explicit,
numerically stable, and easy to implement and because it preserves several
geometric properties of the exact evolution operator, such as unitarity,
symplecticity, and time reversibility.

\section{Numerical examples}

In the following, we use natural units with $m=\hbar=k_{B}=\hat{\mu}=1$. As in
Ref.~\onlinecite{Begusic_Vanicek:2020}, the excited-state surface is a
one-dimensional Morse potential,
\begin{equation}
V_{e}(q)=V_{e}(q_{\text{ref}})+\frac{\omega_{e}}{4\chi}\left[  1-e^{-\sqrt
{2m\omega_{e}\chi}(q-q_{\text{ref}})}\right]  ^{2},
\end{equation}
with the equilibrium position $q_{\text{ref}}=1.5$, the minimum energy
$V_{e}(q_{\text{ref}})=10$, the fundamental frequency $\omega_{e}=0.9$ at
$q_{\text{ref}}$, and the anharmonicity measure $\chi=0.02$. The initial,
ground-state potential energy surface is harmonic and given by
Eq.~(\ref{eq:harmonic}) with $q_{\text{ref}}^{\prime}=0$ and $\kappa=1$.

The spectra are computed either by Boltzmann averaging [Eq.~(\ref{eq:boltz})] or
with thermofield dynamics [Eq.~(\ref{eq:tfd_ct})] at scaled temperatures
$T_{\omega}=k_{B}T/(\hbar\omega_{g})=0$, $0.5$, $1$, and $2$, where
$\omega_{g}=\sqrt{\kappa/m}$ is the ground-state vibrational frequency. Here,
as a demonstration, we apply the standard second-order TVT algorithm on a grid
with 256 points for each degree of freedom. The wavepacket is propagated for a
total time of 1000 with a time step of 0.1. A Gaussian damping function with a
half-width at half-maximum of 15 is applied to the autocorrelation functions
before the spectra are evaluated with Eq.~(\ref{eq:spec}).

\begin{figure}
[htbp]\centering\includegraphics{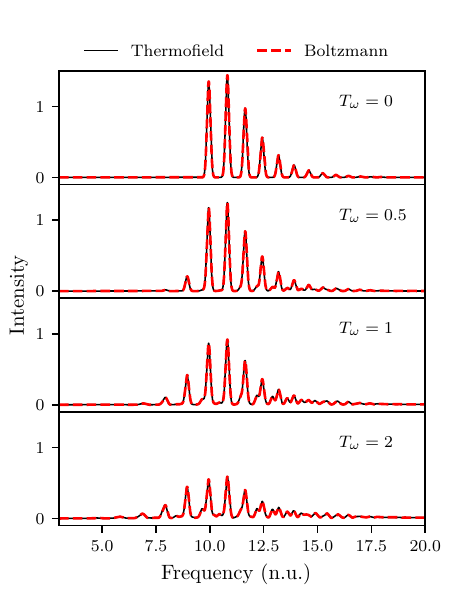}
\caption{Comparison
of spectra obtained via  coherence thermofield dynamics (black solid line) and
Boltzmann averaging (red dashed line) at four different scaled temperatures
$T_{\omega}$.} \label{fig:tfd_vs_boltz}
\end{figure}

\begin{figure}
[htbp]\centering\includegraphics{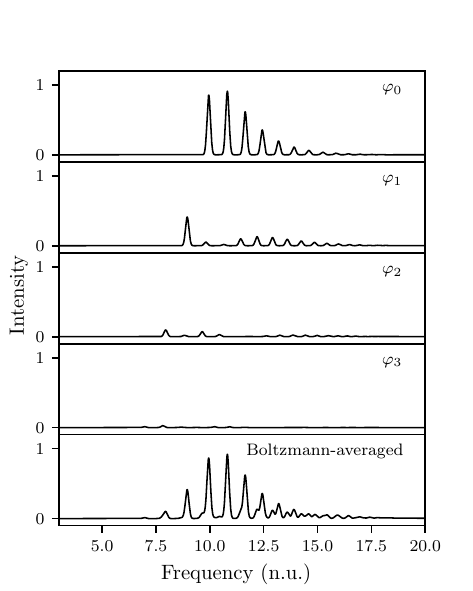}
\caption{Contributions
from different vibrational levels, scaled by their Boltzmann probability of
occupation, to the Boltzmann-averaged spectrum at $T_{\omega}=1$.}
\label{fig:boltz_decomp}
\end{figure}

As shown in Fig.~\ref{fig:tfd_vs_boltz}, the thermofield split-operator
Fourier method gives the same results as the Boltzmann-weighted average. As
the temperature increases, the spectral range broadens and additional peaks
(the \textquotedblleft hot bands\textquotedblright) appear. In
Ref.~\onlinecite{Begusic_Vanicek:2020}, a comparison was made between the
thermal spectra obtained using the thermofield thawed Gaussian approximation and the
\textquotedblleft exact\textquotedblright\ spectra obtained by computing
individual Franck-Condon factors by numerical integration. With the
time-dependent split-operator approach, we avoid the need to preselect the
relevant transitions and can give numerically exact results for arbitrary
global potentials, even if their vibrational eigenfunctions do not have
analytical expressions. This approach thus has the potential to validate
semiclassical methods in a wider range of situations, including those with a
fitted global potential energy surface.

In Fig.~\ref{fig:boltz_decomp}, we show the decomposition of the thermal
spectrum at $T_{\omega}=1$ into the contributions from different initial
vibrational levels. Only spectra from levels $n=0$, $1$, $2$, and $3$ are
shown since the contributions from $n>3$ are negligible here (the Boltzmann
weight is 0.0002 for $n=4$). As expected, the higher vibrational levels are
the source of hot bands and expand the spectral range. Whereas the Boltzmann
averaging requires propagating each initial vibrational level separately, a
single thermofield wavepacket trajectory takes into account all thermal contributions.

\begin{figure}
\centering\includegraphics{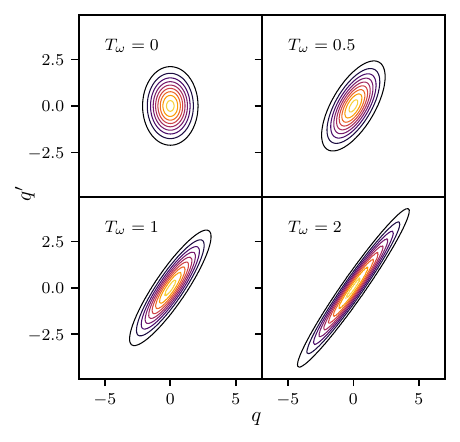}
\caption{Initial thermofield wavepackets at
different scaled temperatures $T_{\omega}$; $q$ is the physical degree of freedom, whereas
$q^{\prime}$ is fictitious.} \label{fig:wp_temps}
\end{figure}

With a grid-based method, we can easily visualize the wavepacket and its
evolution. Figure~\ref{fig:wp_temps} shows the initial thermofield wavepacket
at different temperatures in the augmented coordinate space. As the
temperature increases, the off-diagonal block $\mathrm{B}$ of the initial
width matrix~(\ref{eq:tfd_width}) increases and the initial thermofield
wavepacket rotates and stretches as a result.

\begin{figure}
\centering\includegraphics{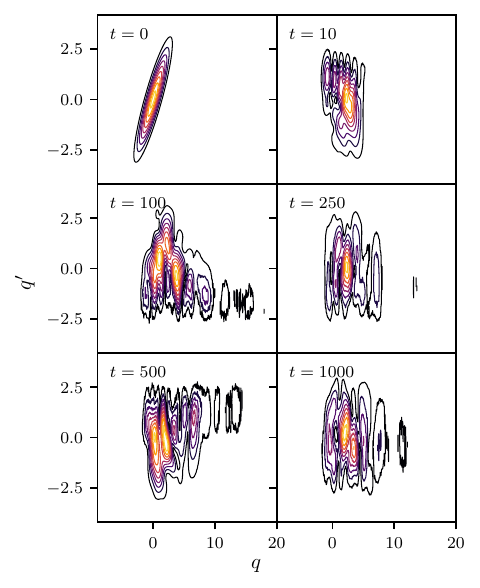}
\caption{Evolution of the
thermofield wavepacket (for $T_{\omega}=1$) in the augmented configuration space
at different times $t$; $q$ is the physical degree of freedom, whereas
$q^{\prime}$ is fictitious.} \label{fig:wp_evolution}
\end{figure}

In Fig.~\ref{fig:wp_evolution}, we show the evolution of the wavepacket at
$T_{\omega}=1$. Whereas the extent of the wavepacket over the fictitious
degree of freedom $q^{\prime}$ remains approximately constant, the initial
excitation of the coordinate $q$ on the anharmonic excited-state surface leads
to a significant evolution of the wavepacket in the physical dimension.

\section{Discussion and conclusion}

We have described a very simple method for computing vibronic spectra at
finite temperatures. This method applies the Fourier split-operator propagation
algorithm to the thermofield wavepacket representation of the electronic
coherence, avoiding both Boltzmann averaging and solving the von Neumann
equation. Instead, we solve the zero-temperature Schr\"{o}dinger equation in
an augmented space. The method yields quantum mechanically exact spectra of
arbitrary global potentials, making it a valuable tool for validating
semiclassical and other approximate techniques for finite-temperature spectra
calculations. In contrast to Boltzmann averaging spectral
contributions from different initial vibrational levels, the thermofield
dynamics approach obtains the exact result for a given temperature from a
single propagation. At high temperatures, as the number of contributing states
increases, the thermofield approach becomes more favorable. If highly
accurate results are desired, the simple approach described here can be made
both more accurate and efficient by implementing high-order geometric
integrators obtained by symmetric composition of the elementary second-order
algorithm, in the same way as for the zero-temperature split-operator
calculations.\cite{Wehrle_Vanicek:2011,Choi_Vanicek:2019,Roulet_Vanicek:2019,Roulet_Vanicek:2021}%

The method presented here is based on the full tensor-product grid and,
therefore, is limited to systems with very few degrees of freedom.\ To extend
its applicability to higher-dimensional systems, one can combine the
split-operator thermofield dynamics with adaptive moving
grids,\cite{book_Thompson_Mastin:1997,Pettey_Wyatt:2006,Choi_Vanicek:2019a}
matching-pursuit coherent-state basis,\cite{Chen_Batista:2006} tensor-train
representation,\cite{Greene_Batista:2017} or any other technique designed for
the zero-temperature case. In fact, tensor-train representations have already
been used extensively\cite{Borrelli_Gelin:2016,Borrelli_Gelin:2021} for the
thermofield dynamics, although not in conjunction with the split-operator
Fourier method. Here, we have intentionally restrained ourselves to the
simplest, full-grid method, because we did not want to obscure the main idea
by combining it with various strategies for increasing the accessible
dimensionality. We believe that even the full-grid version can be interesting
for researchers who would like to quickly check the thermal effects on their
zero-temperature spectra of low-dimensional systems without implementing new codes.

The method is not restricted to linear spectra. In the same way as the
Gaussian,\cite{Begusic_Vanicek:2021a} multiconfigurational
Ehrenfest,\cite{Chen_Gelin:2021} or optimized
mean-trajectory\cite{Polley_Loring:2022} semiclassical thermofield dynamics,
the method can be applied to more sophisticated experimental techniques, such
as pump-probe and two-dimensional spectroscopies. Beyond vibronic spectra,
this approach may also find applications in other finite-temperature dynamics
problems, such as the calculation of the rates of internal
conversion.\cite{book_Heller:2018,Fischer_Saalfrank:2021,Wang_Shuai:2021,Wenzel_Mitric:2023}%

\begin{acknowledgments}
The authors thank Tomislav Begu\v{s}i\'{c} for discussions and acknowledge the
financial support from the EPFL.
\end{acknowledgments}

\section*{Author declarations}

\subsection*{Conflict of interest}

The authors have no conflicts to disclose.

\section*{Data availability}

The data that support the findings of this study are available within the article.

\appendix

\section*{Appendix: Derivation of Eq.~(\ref{eq:tfd_ct})}

The correlation function (\ref{eq:thermal_ct}) can be re-expressed
as\cite{Begusic_Vanicek:2020}
\begin{align}
C(t)  &  =\operatorname{Tr}(\hat{\rho}^{1/2}\hat{\mu}^{\dagger}e^{-i\hat
{H}_{e}t/\hbar}\hat{\mu}\hat{\rho}^{1/2}e^{i\hat{H}_{g}t/\hbar})\nonumber\\
&  =\int dqdq^{\prime}\langle q^{\prime}|\hat{\rho}^{1/2}\hat{\mu}^{\dagger
}|q\rangle\langle q|e^{-i\hat{H}_{e}t/\hbar}\hat{\mu}\hat{\rho}^{1/2}%
e^{i\hat{H}_{g}t/\hbar}|q^{\prime}\rangle\nonumber\\
&  =\int d\bar{q}\bar{\psi}_{0}(\bar{q})^{\ast}\bar{\psi}_{t}(\bar{q}),
\label{eq:C_t_7}%
\end{align}
where $\bar{\psi}_{t}(\bar{q})$ is the propagated thermofield wavefunction,
\begin{align}
\bar{\psi}_{t}(\bar{q})  &  =\langle q|e^{-i\hat{H}_{e}t/\hbar}\hat{\mu}%
\hat{\rho}^{1/2}e^{i\hat{H}_{g}t/\hbar}|q^{\prime}\rangle\nonumber\\
&  =e^{-i\hat{H}_{e}(q)t/\hbar}e^{i\hat{H}_{g}(q^{\prime})t/\hbar}\langle
q|\hat{\mu}\hat{\rho}^{1/2}|q^{\prime}\rangle\nonumber\\
&  =e^{-i\bar{H}(\bar{q})t/\hbar}\bar{\psi}_{0}(\bar{q}), \label{eq:C_t_6}%
\end{align}
$\bar{q}=(q,q^{\prime})^{T}$ is a $2D$-dimensional coordinate vector, and
$\bar{H}(\bar{q})$ is the augmented Hamiltonian~(\ref{eq:tfd_hamil}).

\bibliographystyle{aipnum4-2}
\bibliography{tfd_qm_v08}

\end{document}